\documentclass[twocolumn,showpacs,aps]{revtex4}

\usepackage{graphicx}
\usepackage{latexsym}
\usepackage{amssymb}
\usepackage{color}
\usepackage[center]{subfigure}

\begin{document}

\newcommand{\bq}{\begin{equation}}
\newcommand{\eq}{\end{equation}}
\newcommand{\bqn}{\begin{eqnarray}}
\newcommand{\eqn}{\end{eqnarray}}
\newcommand{\nb}{\nonumber}
\newcommand{\lb}{\label}
\newcommand{\noi}{\noindent}

\title{{A Dynamical Solution in Ho\v{r}ava-Lifshitz Gravity in the IR Limit}}

\author{O. Goldoni $^{1}$}
\email{otaviosama@gmail.com}
\author{M. F. A. da Silva $^{1}$}
\email{mfasnic@gmail.com}
\author{R. Chan $^{2}$}
\email{chan@on.br}
\author{V. H. Satheeshkumar$^{1, 2}$}
\email{vhsatheeshkumar@gmail.com}
\author{J. F. Villas da Rocha $^3$}
\email{jfvroch@pq.cnpq.br}

\affiliation{
$^{1}$Departamento de F\'{\i}sica Te\'orica,
Instituto de F\'{\i}sica, Universidade do Estado do Rio de Janeiro,
Rua S\~ao Francisco Xavier 524, Maracan\~a
20550-900, Rio de Janeiro, RJ, Brasil.\\
$^{2}$Coordena\c{c}\~ao de Astronomia e Astrof\'{\i}sica, 
Observat\'orio Nacional, Rua General Jos\'e Cristino, 77, S\~ao Crist\'ov\~ao  
20921-400, Rio de Janeiro, RJ, Brazil.\\
$^{3}$Universidade Federal do Estado do Rio de Janeiro,
Instituto de Bioci\^encias,
Departamento de Ci\^encias Naturais, Av. Pasteur 458, Urca,
CEP 22290-240, Rio de Janeiro, RJ, Brazil.} 

\date{\today}

\begin{abstract}
Non-stationary null dust in a spherically symmetric spacetime is studied in the context of a general-covariant Ho\v{r}ava-Lifshitz theory. The non-minimal coupling to matter is considered in the infrared limit. The aim of this paper is to study whether the collapse of a null dust-like fluid can be a solution of Ho\v{r}ava-Lifshitz theory in the infrared limit. We have shown that the unique possible solution is static. 
This solution represents a Minkowski spacetime since the energy density is null. 

\end{abstract}


\maketitle

\section{Introduction}

Finding a consistent theory of quantum gravity is still one of the fundamental problems in
contemporary theoretical physics. This problem is further complicated by the fact that we have not yet observed any quantum effects of gravitation.

As it has been shown that General Relativity (GR) is not perturbatively renormalizable, one has to find new strategies \cite{Satheeshkumar:2015aga} \cite{Isham:2002ws}.

This is the motivation for the recent quantum theory of gravity with 
anisotropic scaling proposed by Petr Ho{\v r}ava \cite{Horava2009}, the so-called Ho{\v r}ava-Lifshitz (HL) theory.

Because of its power-counting renormalizability,  HL theory is considered to be one of the promising approaches to quantum gravity. {Recently, the
renormalization of the  projectable HL theory without the extra
U(1) symmetry was proved by Barvinsky at al. \cite{Barvinsky2016}}.
The existence of anisotropic scaling implies that there is a preferred time-coordinate \cite{Visser:2011mf}. This helps in slicing the four dimensional manifold into foliations. Thus, it is natural to use the ADM decomposition of the metric \cite{ADM}. 

There is an extensive literature on HL gravity, where some problems of internal consistency and compatibility with observations are discussed.
References \cite{Visser:2011mf}, \cite{Weinfurtner}, \cite{Sotiriou} and \cite{Wang:2017brl} review and discuss these issues. 
Problems including strong coupling, breaking of unitarity and other inconsistencies are addressed
in references \cite{Charmousis} \cite{Blas} \cite{Henneaux}. 
It is important to emphasize that, although the motivation for the construction of this theory has
been the possibility of quantization of gravitation, this has not been done yet in four dimensions \cite{BaoFei}.

The gravitational collapse is a well explored phenomenon {in GR}, for this reason it is
interesting to study this in the context of HL theory.  Also, it becomes a compatibility test as HL theory should converge to GR in the IR. 
Recent literature on gravitational collapse in HL theory has shown that very little effort has been made to understand the theory even classically {\cite{Grennwald2013}}. Some of the previous works published
by three of the authors have attempted to fill the gap 
\cite{Goldoni2014} \cite{Goldoni2015} \cite{Goldoni2016}.
The present paper is a continuation of our previous papers to understand the HL theory in the IR.

Our aim in the present work is to revisit the null dust solution in the projectable case of HL theory, 
but without using the Vaidya's metric. We have chosen a null dust fluid because it is the simplest null fluid that we can have.  Besides, we would like to compare the results with our previous work \cite{Goldoni2016} with non-null dust fluid. Our original proposal was to use the most general metric without any hypothesis to simplify the equations, but we found that these equations cannot be solved analytically. So, 
by placing some conditions on the metric, we have succeed in finding a particular solution. 
One of these conditions is that the variables of the metric are separable.
{ This consideration is strong and it excludes Vaidya's metric as a solution { in GR}. However, as it was shown by some of us \cite{Goldoni2014}\cite{Goldoni2015}, Vaidya's metric is not a solution {of HL} in the IR. 
That is, the same fluid or metric, does not necessarily correspond to the same solution in both theories, HL and GR. Therefore, the hypothesis that there exits some solution for a null fluid with variable separation in HL in the IR limit is plausible.}

The paper is organized as follows. In Section II, we present a brief introduction to HL theory with coupling with matter {\cite{Lin2013}\cite{Lin2014}}. In Section III, we study the null dust solution in the infrared limit. In Section IV, we discuss the results.

\section{General Covariant Ho\v{r}ava-Lifshitz Gravity with Coupling with the Matter}

We present a summary of the HL gravity with the non-minimal coupling to matter. 
For more details, we refer the readers to  {\cite{Lin2012}\cite{Lin2013} \cite{Lin2014}} \cite{ZWWS} \cite{ZSWW}.

A line element in the Arnowitt-Deser-Misner (ADM) form is given by \cite{ADM},
\bqn
ds^{2} &=& - N^{2}dt^{2} + g_{ij}\left(dx^{i} + N^{i}dt\right)
\left(dx^{j} + N^{j}dt\right), \nb\\
& & ~~~~~~~~~~~~~~~~~~~~~~~~~~~~~~  (i, \; j = 1, 2, 3),
\lb{ds2}
\eqn
where the non-projectability condition implies that the lapse is a function of spacetime, $N \equiv N(t,x^i)$.
{However in this work, since we use the non-minimal coupling, we must have $N=1$,
i.e., we will work with a projectable metric.} 

In reference {\cite{Lin2013}\cite{Lin2014}}, Lin \textit{et al.} proposed that, in the IR, it is possible to have matter fields universally coupled to the ADM components through the transformations
\bqn
\lb{eq8-1}
& & \tilde{N} = F N,\;\;
\tilde{N}^i = N^i + Ng^{ij} \nabla_j\varphi,\nb\\
\;\; \tilde{g}_{ij} = \Omega^2g_{ij},
\eqn
with
\bqn
\lb{eq8-1a}
& & F = 1 - a_1\sigma, \;\;\;
 \Omega = 1 - a_2\sigma,
\eqn
where
\bqn
\sigma &\equiv &\frac{A - {\cal{A}}}{N},\nb\\
{\cal{A}} &\equiv& - \dot{\varphi}  + N^i\nabla_i\varphi
+\frac{1}{2}N\left(\nabla^i\varphi\right)\left(\nabla_i\varphi\right),\nb\\
\eqn
and where $A$ and $\varphi$ 
are the gauge field and the Newtonian prepotential,
respectively, 
and $a_1$ and $a_2$ are two arbitrary coupling constants. Note that by
setting the first terms in $F$ and $\Omega$ to unity, we have used the
freedom to rescale the units of time and space. We also have
\bq
\lb{eq8-2}
\tilde{N}_i =\Omega^2\left(N_i + N\nabla_i\varphi\right),\;\;\;
\tilde{g}^{ij} = \Omega^{-2}g^{ij}.
\eq
In terms of these newly defined quantities, the matter action can be written as
\bqn
\lb{eu7}
S_{m} &=& \int{dtd^3x \tilde{N}\sqrt{\tilde{g}}\;  \tilde{\cal{L}}_{m}\left(\tilde{N}, \tilde{N}_i, \tilde{g}_{ij}; \psi_{n}\right)},
\eqn
where $\psi_n$ collectively  stands for all matter fields. One can then define the 
matter stress-energy in the ADM decomposition, with the non-minimal coupling. The different components are given by (for the details see {\cite{Lin2012}\cite{Lin2013}\cite{Lin2014}})
\bqn
\lb{Tab}
\rho_H (GR) & = & T_{\mu\nu}n^\mu n^\nu \equiv J^t =  -\frac{\delta(\tilde{N}\tilde{\cal{L}}_{m})}{\delta(\tilde{N})}\nb\\
S^i (GR) & = & -T_{\mu\nu} h^{(4)i\mu}n^\nu \equiv J^i =  \frac{\delta(\tilde{N}\tilde{\cal{L}}_{m})}{\delta(\tilde{N}_i)}\nb\\
S^{ij} (GR) & = & T_{\mu\nu}h^{(4)i\mu}h^{(4)j\nu} \equiv \nb\ \\
& & \tau^{ij} = \frac{2}{\tilde{N}\sqrt{\tilde{g}}}\frac{\delta(\tilde{N}\sqrt{\tilde{g}}\tilde{\cal{L}}_{m})}{\delta(\tilde{g}_{ij})},
\eqn
where ${h^{(4)}}_{\mu\nu}$ is the projection operator, defined as ${h^{(4)}}_{\mu\nu}\equiv {g^{(4)}}_{\mu\nu}+n_\mu n_\nu$ and $n^\mu$ is the normal vector to the hypersurface $t=$ constant, defined as $n^\mu = \frac{1}{N}(-1,N^i)$. 

The total action of the theory can be written as 
\bqn 
\lb{TA}
S &=& \zeta^2\int dt d^{3}x  \sqrt{g}N \Big({\cal{L}}_{K} -
{\cal{L}}_{{V}} +  {\cal{L}}_{{A}}+ {\cal{L}}_{{\varphi}}  + {\cal{L}}_{S}+\nb\\
& &\frac{1}{\zeta^2} {\cal{L}}_{M}\Big), 
\eqn
where $g={\rm det}(g_{ij})$, $N$ is given in the equation (\ref{ds2}).

The Ricci and Riemann tensors 
$R_{ij}$ and $R^{i}_{\;\; jkl}$  all refer to the 3-metric $g_{ij}$, with 
$R_{ij} = R^{k}_{\;\;ikj}$ and
\bqn \lb{2.6}
 R_{ijkl} &=& g_{ik}R_{jl}   +  g_{jl}R_{ik}  -  g_{jk}R_{il}  -  g_{il}R_{jk}\nb\\
 &&    - \frac{1}{2}\left(g_{ik}g_{jl} - g_{il}g_{jk}\right)R,\nb\\
K_{ij} &\equiv& \frac{1}{2N}\left(- \dot{g}_{ij} + \nabla_{i}N_{j} +
\nabla_{j}N_{i}\right),\nb\\
{\cal{G}}_{ij} &\equiv& R_{ij} - \frac{1}{2}g_{ij}R,\nb\\
a_{i} &\equiv& \frac{N_{,i}}{N},\;\;\; a_{ij} \equiv \nabla_{j} a_{i},
\eqn
where $N_i$ is defined in the ADM form of the
metric \cite{ADM}, given by equation (\ref{ds2}).

The variations of the action $S$ given by equation (\ref{TA}) with respect to $N$ and $N^{i}$ 
give rise to the Hamiltonian and momentum constraints,
\bqn \label{hami}
{\cal{L}}_K + {\cal{L}}_V + F_V-F_\varphi-F_\lambda+{\cal{H}}_S= 8 \pi G J^t,\;\;
\eqn
\bqn \label{mom}
&& M_S^i+\nabla_j \bigg\{\pi^{ij} -(1-\lambda)g^{ij}\big(\nabla^2\varphi+a_k\nabla^k\varphi\big)\nb\\
&& ~~~~~~~~~~~~~~~ - \varphi {\cal{G}}^{ij} - \hat{{\cal{G}}}^{ijkl} a_l \nabla_k \varphi\bigg\} = 8\pi G J^i, ~~~~
\lb{jui}
\eqn
where
\bqn
{\cal H}_S&=&\frac{2\sigma_1}{N}\nabla_i\left[a^i\left(A-{\cal A}\right)\right]-\frac{\sigma_2}{N}\nabla^2\left(A-{\cal A}\right)\nb\\          
&&+\frac{1}{2} a_S \nabla_j\varphi\nabla^j\varphi,\nb\\       
M_S^i&=&-\frac{1}{2}a_S\nabla^i\varphi, \nb\\                                    
J^i &=& -N \frac{\delta {\cal{L}}_M}{\delta N_i},\;\;
J^t = 2 \frac{\delta (N {\cal{L}}_M)}{\delta N},\nb\\
\pi^{ij}&=&-K^{ij}+\lambda K g^{ij},
\eqn
with
\bqn
a_S=\sigma_1a_ia^i+\sigma_2a^i_i,
\eqn
and  $F_V$, $F_{\varphi}$ and $F_\lambda$ are given in the Appendix A of the reference \cite{Goldoni2016}.

Variations of action $S$ given by equation (\ref{TA}) with respect to $\varphi$ and $A$ yield, respectively,
\bqn \label{phi}
&& \frac{1}{2} {\cal{G}}^{ij} ( 2K_{ij} + \nabla_i\nabla_j\varphi  +a_{(i}\nabla_{j)}\varphi)\nb\\
&& + \frac{1}{2N} \bigg\{ {\cal{G}}^{ij} \nabla_j\nabla_i(N \varphi) - {\cal{G}}^{ij} \nabla_j ( N \varphi a_i)\bigg\}\nb\\
&& - \frac{1}{N} \hat{{\cal{G}}}^{ijkl} \bigg \{ \nabla_{(k} ( a_{l)} N K_{ij}) + \frac{2}{3} \nabla_{(k} (a_{l)} N \nabla_i \nabla_j \varphi)\nb\\
&& - \frac{2}{3} \nabla_{(j} \nabla_{i)} (N a_{(l} \nabla_{k)} \varphi) + \frac{5}{3} \nabla_j (N a_i a_k \nabla_l \varphi)\nb\\
&& + \frac{2}{3} \nabla_j (N a_{ik} \nabla_l \varphi)\bigg\}+\Sigma_S \nb\\
&& + \frac{1-\lambda}{N} \bigg\{\nabla^2  \left.[N (\nabla^2 \varphi + a_k \nabla^k \varphi)\right.] \nb\\
&& - \nabla^i [N(\nabla^2 \varphi + a_k \nabla^k \varphi) a_i] \nb\\
&&+ \nabla^2 (N K) - \nabla^i ( N K a_i)\bigg \}
 = 8 \pi G J_\varphi,
\eqn
where,
\bqn \label{Sigma}
\Sigma_S & = & -\frac{1}{2N}\Bigg\{\frac{1}{\sqrt{g}}\frac{\partial}{\partial t}\left[\sqrt{g}a_S\right]\nb\\
&&-\nabla_k\left[\left(N^k+N\nabla^k\varphi\right)a_S\right]\Bigg\}
\eqn
and
\bqn \label{ja}
R-a_S= 8 \pi G J_A,
\eqn
where
\bqn
J_\varphi = -\frac{\delta {\cal{L}}_M}{\delta \varphi},\;\;\;
J_A= 2 \frac{\delta ( N {\cal{L}}_M)}{\delta A}.
\eqn

The variation of the action $S$ given by equation (\ref{TA}) with respect to $g_{ij}$ give us the 
dynamical equations,
\bqn \label{dyn}
\frac{1}{\sqrt{g}N} \frac{\partial}{\partial t}\left(\sqrt{g} \pi^{ij}\right)+2(K^{ik}K^j_k-\lambda K K^{ij})\nb\\
-\frac{1}{2}g^{ij}{\cal{L}}_K+\frac{1}{N}\nabla_k (\pi^{ik}N^j+\pi^{kj}N^i-\pi^{ij}N^k)\nb\\
+F^{ij}-F^{ij}_S-\frac{1}{2}g^{ij}{\cal{L}}_S+F^{ij}_a-\frac{1}{2}g^{ij}{\cal{L}}_A+F^{ij}_\varphi\nb\\
-\frac{1}{N}(AR^{ij}+g^{ij}\nabla^2A-\nabla^j\nabla^iA)
=8\pi G \tau^{ij},\;\;\;\;\;\;
\eqn
where
\bqn
\lb{tauij}
\tau^{ij}&=&\frac{2}{\sqrt{g}N} \frac{\delta(\sqrt{g}N{\cal{L}}_M)}{\delta g_{ij}}, \nb\\
\eqn
and $F^{ij}$, $F^{ij}_S$, $F^{ij}_a$ and $F^{ij}_\varphi$ are given in the 
Appendix A of the reference \cite{Goldoni2016}.

From reference {\cite{Lin2013}\cite{Lin2014}}, we have that
\bqn
\lb{eq12}
\tilde{N} &=&\tilde{N}(N, N_i, g_{ij}, A, \varphi),\nb\\
\tilde{N}_i &=&\tilde{N}_i(N, N_i, g_{ij}, A, \varphi),\nb\\
\tilde{g}_{ij} &=&\tilde{g}_{ij}(N, N_i, g_{ij}, A, \varphi).
\eqn
Thus,
\bqn                                                                           
\lb{eq14}                                                                      
J^{t} &=& 2\Omega^{3}\Bigg\{- \rho  \frac{\delta\tilde{N}}{\delta{N}}+  \frac{\delta\tilde{N}_i}{\delta{N}}S^i   + \frac{1}{2}\tilde{N}  \frac{\delta\tilde{g}_{ij}}{\delta{N}}S^{ij}\Bigg\}. ~~~~~~~~~                            \eqn                                                                         
Similarly, it can be shown that                                                
\bqn                                                                           
\lb{eq15}                                                                      
J^{i} &=& - \Omega^{3}\Bigg\{- \rho  \frac{\delta\tilde{N}}{\delta{N}_i}                                                                            +  \frac{\delta\tilde{N}_k}{\delta{N}_i}S^k   + \frac{1}{2}\tilde{N}  \frac{\delta\tilde{g}_{kl}}{\delta{N}_i}S^{kl}\Bigg\},\nb\\                             
\tau^{ij} &=& \frac{2 \Omega^{3}}{N}\Bigg\{- \rho \frac{\delta\tilde{N}}{\delta{g}_{ij}}                                         +  \frac{\delta\tilde{N}_k}{\delta{g}_{ij}}S^k   + \frac{1}{2}\tilde{N}  \frac{\delta\tilde{g}_{kl}}{\delta{g}_{ij}}S^{kl}\Bigg\},\nb\\
J_{A} &=& 2 \Omega^{3}\Bigg\{- \rho
\frac{\delta\tilde{N}}{\delta{A}}
+  \frac{\delta\tilde{N}_k}{\delta{A}}S^k   + \frac{1}{2}\tilde{N}  \frac{\delta\tilde{g}_{kl}}{\delta{A}}S^{kl}\Bigg\},\nb\\
J_{\varphi} &=& - \frac{1}{N}\Bigg\{\frac{1}{\sqrt{g}}\left(B \sqrt{g}\right)_{,t} - \nabla_{i}\Big[B\left(N^i + N \nabla^i\varphi\right)\Big]\nb\\
&& ~~~~~~~~~ - \nabla_i\left(N\Omega^5 S^i\right)\Bigg\},
\eqn
where
\bqn
\lb{eq16}
B &\equiv& - \Omega^{3}\Bigg\{a_1\rho  - \frac{2a_2\left(1- a_2\sigma\right)}{N}S^k\left(N_k + N\nabla_k\varphi\right)\nb\\
&& -  a_2\left(1- a_1\sigma\right)\left(1- a_2\sigma\right)g_{ij}S^{ij}\Bigg\}.
\eqn

Note that, if we use the projectable case of HL then
${\cal{L}}_{S}=0$, ${M}^i_{S}=0$, ${\cal{H}}_{S}=0$, $\Sigma_{S}=0$ and $a_{S}=0$.
In order to have these quantities in HL, it is
only necessary the non-projectability condition {\cite{Lin2013}\cite{Lin2014}}.

\section{Collapse of a Null Fluid}

For a general spherically spacetime filled with dust
{in GR} \cite{BS09}, the metric can be written as
\bq
ds^2=-dt^2+e^{2\Phi(r,t)} dr^2 + {\cal{R}}(r,t)^2 (d\theta^2 + \sin^2 \theta\; d\phi^2).
\lb{ds2a}
\eq
Note also that this metric is a
projectable one, thus we will use the projectable case of HL, now on.

The energy-momentum tensor for a null fluid is given by
\bq
T_{\mu\nu}=\rho l_\mu l_\nu,
\lb{tmn}
\eq
where the null vector $l_\mu$ is given by
\bq
l_\mu=\delta_\mu^t + e^{\Phi} \delta_\mu^r.
\eq

The projection tensor $h_{\mu\nu}$ is defined as
\bq
h_{\mu\nu}=g_{\mu\nu}+n_\mu n_\nu,
\lb{hmn}
\eq
where $n_\mu=\delta_\mu^t$ thus, we have
\bqn
h_t^t&=&0,\nb\\
h_t^r&=&0,\nb\\
h_r^t&=&0,\nb\\
h_r^r&=&1,\nb\\
h_\theta^\theta &=& h_\phi^\phi =1.
\lb{hmn1}
\eqn

Let us now analyze the case where it exists a non-minimal coupling with matter
{\cite{Lin2012}\cite{Lin2013}\cite{Lin2014}}.  In this case we have the following conditions,
\bqn
\gamma_1 &=& -1,\\ 
a_1 &=& 1,\\
a_2 &=& 0,\\
F &=& 1-A,\\
\Omega &=& 1,\\
\tilde N &=& 1-A,\\
\tilde N^i &=& N^i,\\
\tilde g_{ij} &=& g_{ij},\\
{\cal A} &=& 0,\\
\sigma &=& A,\\
\eqn

Besides, from equations (\ref{eq14})-(\ref{eq16}) we have
\bq
J^t= -2\rho,
\lb{Jt}
\eq
\bq
J^i=2\rho e^{-\Phi} \delta^i_r,
\lb{Jr}
\eq 
\bq
\tau^{ij}=(1-A)\rho e^{-2\Phi} \delta^i_r \delta^j_r,
\lb{taurr}
\eq
\bq
\lb{JA}
J_A = 2\rho,
\eq
\bq
J_\varphi=\dot \rho +  \rho \dot \Phi + 2 \rho \frac{\dot R}{R}-2e^{\Phi}(\rho'+\rho \Phi'),
\lb{Jphi}
\eq 
where we have substituted $B=-\rho$.

In order to be consistent with observations in the infrared limit {\cite{Lin2013}\cite{Lin2014}}, we assume that
\bqn
\beta_1=\beta_2=\beta_3=\beta_4=\beta_5=\beta_6=\beta_7=\beta_8=\beta_9=0,\nb
\eqn
\bqn
\gamma_0=\gamma_2=\gamma_3=\gamma_4=\gamma_5=\gamma_6=\gamma_7=\gamma_8=\gamma_9=0.\nb
\eqn
Thus, we have the vanishing of the cosmological constant, as follows
\bqn
\Lambda_g \equiv \frac{1}{2} \zeta^{2}\gamma_{0}=0.\nb
\eqn

Let us impose that the metric variables are separable.
{ As commented} in the introduction of this paper, 
{these conditions are imposed in order} to be able to solve the equations analytically { using the Maple software.
Hereinafter, the prime and dot symbols denote the partial differentiation with respect to the
coordinate $r$ and $t$, respectively. Thus,} we can write
\bqn
\Phi(r,t)=\log\left(F(t) G(r)\right),
\eqn
and
\bqn
{\cal{R}}(r,t)=S(t) { H(r)}.
\lb{ds2sf}
\eqn

Combining equations (\ref{Jt})-(\ref{Jphi}) and (\ref{2.6})-(\ref{tauij}) we have
\bqn
\rho&=&2 \frac{G' H'}{G^3 F^2 H}-2 \frac{H''}{F^2 G^2 H}-\frac{H'^2}{F^2 G^2 H^2}+\frac{1}{S^2 H^2}.\nb\\
\lb{JA1}
\eqn
Using the equation (\ref{JA1}) and the equations (\ref{Tab})-(\ref{eq16}) we have that
\bqn
&&-2 \frac{H' \dot S}{S H}+2 \frac{H' \dot F}{F H}+4 \frac{G' H'}{G^5 F^4 H}-\frac{4 H''}{G^4 F^4 H}-\nb \\
&&2 \frac{H'^2}{G^4 F^4 H^2}+2\frac{1}{G^2 F^2 H^2 S^2}=0,
\lb{Jr1}
\eqn
\bqn
&&\frac{\dot F^2}{F^2}-\lambda \frac{\dot F^2}{F^2}+2 \frac{\dot F}{F H^2  S^2}-4 \frac{\dot F \lambda}{F H^2 S^2}-4 \lambda \frac{\dot S \dot F}{F  S}=0,\nb\\
\lb{Jt1}
\eqn
\bqn
&&-2 \frac{\dot S G' H'}{G^3 F^2  S H}+3 \frac{\dot S H''}{G^2 F^2  S H}-\lambda \frac{H'' \dot S}{G^2 F^2  S H}-\nb\\
&&2 \frac{\dot F H'^2}{G^2 F^3 H^2}+2 \frac{H'^2 \dot S}{G^2 F^2  S H^2}+2 \frac{\dot F G' H'}{G^3 F^3 H}-\nb\\
&&2 \frac{\dot F H''}{G^2 F^3 H}+4 \frac{G'' H'}{G^2 F H}-2 \frac{G' H'^2}{G^2 F H^2}-4 \frac{H'''}{G F H}+\nb\\
&&2 \frac{F G'}{S^2 H^2}-8 \frac{G'^2 H'}{G^3 F H}+8 \frac{G' H''}{G^2 F H}+4 \frac{H'^3}{G F H^3}-\nb\\
&&4 \frac{G F H'}{S^2 H^3}=0.
\lb{Jphi1}
\eqn

Finally, from the dynamical equations (\ref{dyn}), combined with the second
equation of (\ref{eq15}), we get
\bqn
&&D^{rr}-\tau^{rr}=\nb\\
&&-\frac{1}{2}\frac{1}{G^5 F^4  S^2 H^2}\times\nb\\
&&(-2 G^3  S^2 H^2 \ddot F F+2 G^3  S^2 H^2 \dot F^2-G^3 \dot F^2-\nb\\
&&4 G^3 \dot F \dot S H^2  S F+2 G^3 \lambda  S^2 H^2 \ddot F F-2 G^3 \lambda  S^2 H^2 \dot F^2+\nb\\
&&\lambda G^3 \dot F^2+4 \lambda \dot S \dot F G^3 F H^2  S+4 G A'  S^2 H' H-\nb\\
&&4 H'^2 G  S^2+4 G^3 F^2+4 G A  S^2 H'^2-4 A F^2 G^3+\nb\\
&&4 G^3 F^2 \ddot S H^2 \lambda  S+2 G^3 F^2 \dot S^2 H^2+4 G' H' H  S^2-\nb\\
&&4 H'' G H  S^2-4 A G' H' H  S^2+4 A H'' G H  S^2)=0,\nb\\
\lb{tau11}
\eqn
\bqn
&&D^{\theta\theta}- \tau^{\theta\theta} =\nb\\
&&-\frac{1}{2} \frac{1}{G^3 F^2  S^3 H^3}\times\nb\\
&&(2 A''  S H G-2 A' G'  S H+2 A'  S H' G-2 G'  S H' A+\nb\\
&&2 G'  S H'+2 G^3 \lambda  S H \ddot F F-G^3 \dot F^2 \lambda  S H+G^3 \dot F^2  S H+\nb\\
&&4 F G^3 \dot F  S H' \lambda-2 F G^3 \dot F \dot S H+2  S H'' A G-2  S H'' G+\nb\\
&&4 G^3 F^2 \ddot S H \lambda-2 G^3 F^2 \ddot S H)=0,
\lb{tau22}
\eqn
and
\bq
D^{\phi\phi}= \tau^{\phi\phi}=D^{\theta\theta}\sin^2 \theta.
\eq

Using Maple we have solved equations (\ref{Jr1}), (\ref{Jt1}), (\ref{Jphi1}), (\ref{tau11}) and (\ref{tau22})
simultaneously, given us
\bq
F(t)=c_1,
\lb{ft}
\eq
\bq
S(t)=c_2,
\lb{ht}
\eq
\bq
G(r) = s \frac{\sqrt{( H c_1^2+c_3 c_2^2)  H} H' c_2}{H c_1^2+c_3 c_2^2},
\lb{gr}
\eq
\bq
A(r,t)=1+\frac{\sqrt{H c_1^2+ c_3 c_2^2}F_1(t)}{\sqrt{H}},
\lb{Art}
\eq
where $c_1$, $c_2$, $c_3$ are arbitrary constants, $F_1(t)$ is an arbitrary function of time and $s=\pm 1$.

Substituting equation (\ref{ft}), (\ref{ht}) and (\ref{gr}) into equations (\ref{ds2a}) and (\ref{JA1}) we have finally
\bq
ds^2=-dt^2+s^2\frac{c_1^2HH'^2 c_2^2}{H c_1^2+ c_3 c_2^2} dr^2+c_2^2H^2(d\theta^2 + \sin^2 \theta\; d\phi^2).
\lb{ds2b}
\eq
\bqn
\rho=\frac{s^2-1}{H^2c_2^2s^2}=0,
\lb{rho1}
\eqn
since $s=\pm 1$. Since the energy density is null, this might represent a Minkowski spacetime.

Making a coordinate transformation, such as ${\bf r}=c_2 H$ we obtain
\bq
ds^2=-dt^2+\frac{c_1^2 {\bf r}}{{\bf r} c_1^2+ c_3 c_2^3} d{\bf r}^2+{\bf r}^2(d\theta^2 + \sin^2 \theta\; d\phi^2).
\lb{ds2c}
\eq
Notice that if $c_3\ne 0$ then we have a static non Minkowski spacetime, otherwise we recover the
Minkowski spacetime.  Assuming that
\bq
r_s= -\frac{c_3c_2^2}{c_1^2},
\eq 
we can rewrite equation (\ref{ds2c}) as
\bq
ds^2=-dt^2+\frac{{\bf r}}{{\bf r} - r_s} d{\bf r}^2+{\bf r}^2(d\theta^2 + \sin^2 \theta\; d\phi^2).
\lb{ds2d}
\eq

Thus, the quantity $A({\bf r},t)$ is given by
\bq
A({\bf r},t)=1+c_1 \sqrt{1-\frac{r_s}{{\bf r}}}F_1(t).
\eq

Since we have used a projectable metric assuming a time coordinate transformation
such as $dt'=(1-A)dt$, then we must have $r_s=0$, because $A$ must be time dependent only, i.e., $A=A(t)$.
Thus, the unique possible solution we have is the Minkowski spacetime.

\section{Conclusion}

We have studied the spherically symmetric spacetime filled with a null dust fluid  in the framework of general covariant HL theory with the non-minimal coupling {\cite{Lin2012}\cite{Lin2013}\cite{Lin2014}}, in the infrared limit.  We have analyzed if such a solution can be described in the general covariant HL gravity \cite{ZWWS,ZSWW}.
We have found a solution that is static. This solution represents a Minkowski spacetime in GR, since the energy density is null.

At this point we have some comments to make.  
{ Notice that we have no longer expected to find Vaidya's solution. As shown by some of us \cite{Goldoni2014} \cite{Goldoni2015}, Vaidya's metric is not a solution for a null fluid in the HL in the IR. So, we could ask if there is any other null fluid solution different from the Vaidya's in HL gravity. We have found one with variables separated.}

{ It should be noted that the techniques used here to {find} a solution, that is, separation of variables, was  used before in our previous paper \cite{Goldoni2016}, working with timelike dust fluid. There, we {found interesting} results. {One of the solutions} has the same behavior as in GR but another one has a temporal behavior that can be interpreted 
as a bouncing. Therefore, while timelike fluids retain some analogy between the HL and GR in the IR limit, the same can not be concluded for null fluids.}

\begin{acknowledgements} 

The financial assistance from FAPERJ/UERJ (MFAS, VHS) are gratefully acknowledged.
The author (RC) acknowledges the financial support from FAPERJ 
(no. E-26/171.754/2000, E-26/171.533/2002, E-26/170.951/2006, E-26/110.432/2009 and E-26/111.714/2010). The authors (RC and MFAS) also acknowledge the 
financial support from Conselho Nacional de Desenvolvimento Cient{\'i}fico e
Tecnol{\'o}gico - CNPq - Brazil (no. 450572/2009-9, 301973/2009-1 and 
477268/2010-2). The author (MFAS) also acknowledges the financial support 
from Financiadora de Estudos e Projetos - FINEP - Brazil (Ref. 2399/03).
The authors (OG, VHS) thank the financial support from CAPES/Science without Borders (no. A 045/2013).
It is our pleasure to thank Anzhong Wang for helpful discussions and comments about this work.

\end{acknowledgements}

\end{document}